# Accumulative magnetic switching of ultra-high-density recording media by circularly polarized light

[1]Y.K. Takahashi*, [2]R. Medapalli, [1]S. Kasai, [1]J. Wang, [1]K. Ishioka, [3]S.H. Wee, [3]O. Hellwig, [1]K. Hono and [2]E.E. Fullerton

[1] Magnetic Materials Unit, National Institute for Materials Science, 1-2-1 Sengen, Tsukuba 305-0047, Japan

[2] Center for Memory and Recording Research, University of California San Diego, 9500 Gilman Drive, La Jolla, CA 92093-0401, USA

[3] San Jose Research Center, HGST a Western Digital company, 3403 Yerba Buena Road, San Jose, CA 95135, USA

Corresponding author,

Y.K. Takahashi

Magnetic Materials Unit, National Institute for Materials Science, 1-2-1 Sengen, Tsukuba 305-0047, Japan

e-mail: takahashi.yukiko@nims.go.jp

**Manipulation of the magnetization by external energies other than magnetic field, such as spin-polarized current[1-4], electric voltage[5,6] and circularly polarized light[7-11] gives a paradigm shift in magnetic nanodevices. Magnetization control of ferromagnetic materials only by circularly polarized light has received increasing attention both as a fundamental probe of the interactions of light and magnetism but also for future high-density magnetic recording technologies. Here we show that for granular FePt films, designed for ultrahigh-density recording, the optical magnetic switching by circularly polarized light is an accumulative effect from multiple optical pulses. We further show that deterministic switching of high anisotropy materials by the combination of circularly polarized light and modest external magnetic fields, thus revealing a pathway towards technological implementation.**

Deterministic control of magnetization by light, often referred to as all-optical switching (AOS), is an attractive recording method for magnetic nanotechnologies because magnetization control becomes possible without the need of an external magnetic field[7-11] and therefore incorporates the potential for ultra-fast magnetization switching up to 1000 times faster than that by magnetic fields while using lower energies[12]. The first demonstration of the magnetization switching by light was in ferrimangnetic GdFeCo film which is a magneto-optical material[7] where the Gd and FeCo spin sub-lattices are antiferromagnetically exchange coupled. While several mechanisms for the ultrafast magnetization switching of GdFeCo have been explored[14-16], the current understanding for AOS in GdFeCo is that the ultrafast laser excitation demagnetizes the two sublattices at different time-scales[12] resulting in a transient ferromagnetic state where the Gd and FeCo sublattices are parallel. At later times as the Gd moment reemerges in the antiparallel orientation due to the exchange interaction the sublattice moments end up in the opposite direction as compared to the initial state. The deterministic helicity dependent switching observed in the experiments results from helicity dependent absorption (dichroism) which favors one magnetic configuration over the other for a narrow range of fluence[17].

It was believed that AOS occurs only in ferrimagnetic materials including synthetic structures[7-10] since the mechanism determined for GdFeCo films required antiparallel exchange of two sublattice systems. However, recently Lambert *et al*. reported that the optical control of the magnetization occurs in ferromagnets including Co-based multilayer thin films and FePtAg-C granular thin film materials[11]. Therefore the potential mechanisms for AOS in ferromagnetic materials must be reexamined. In addition the observation of AOS in granular FePt-based films being developed for heat-assisted magnetic recording (HAMR)[18,19] directly impacts the magnetic recording community, since it provides a potential solution to the so-called "trilemma problem" in high-density hard disk drives (HDDs) beyond 1 Tbit/in$^2$ [20,21]. Here, to probe both the mechanisms of optical reversal and its potential for novel technologies, we report the observation of accumulative magnetic switching from multiple circularly polarized light pulses on FePt-C granular HAMR media.

Figure 1(a) shows the magnetic image of a demagnetized FePt-C granular film after scanning it with both right circularly polarized (RCP) or left circularly polarized (LCP) light pulses. As was reported in Ref. 11 the optical pulses induce a net magnetization in the FePt granular media and the sign of the magnetization is determined by the helicity of the light. To quantify the optically-induced magnetization changes we patterned the granular magnetic film into a Hall cross (see the methods section and Fig. S2). The Hall resistance corresponds to the average out-of-plane component in the magnetization of the FePt grains within the Hall-cross area. Hall resistance changes are normalized by the field dependent hysteresis loop, which reveals ~ 2.75 Ω change in Hall resistance between negative and positive saturation in Fig. 1(b). We first sweep the laser over the Hall cross in a similar way as shown in Fig. 1(a) and measure an induced magnetization of roughly 66% of the saturation magnetization as shown in Fig. 1(b). We then fixed the laser over the Hall cross region to measure the evolution of the magnetization to a series of ultrafast laser pulses. We initially used 40 optical pulses for the first two exposure steps, and then, increased to 80 pulses for the next six exposure steps (see methods and Fig. S2). The Hall resistance was measured after each sequence. The initial state is remanence after applying saturating magnetic fields of -7 T (Fig. 1(c)) and 7 T (Fig. 1(d)). Figure 1(c) shows the normalized Hall resistance change after the exposure to RCP, linearly polarized and LCP. For RCP light, the normalized magnetization gradually decreases to zero, then reverses and saturates at about -0.5. This indicates that ~3/4 of the FePt grains switch to the opposite direction. On the other hand, the exposure to left circularly polarized (LCP) light decreases the magnetization to about half of the initial value, corresponding to the switching of ~1/4 the FePt grains. For exposure to linearly polarized light, the normalized Hall resistance gradually approaches zero. In the case of the opposite initial state (negative saturation) shown in Fig. 1(d), RCP, LCP and linearly polarized light exposures result in the same final normalized magnetization of -0.5, 0.5 and zero, respectively. Thus, the magnetization state after exposure to polarized light only depends on the helicity of the light and is independent on the initial magnetization state. More importantly, the results show that the final state is only achieved after ~500 pulses are applied for this specific laser power. The behavior shown in Fig. 1(c) and (d) is generally seen also for other pulse sequences and laser powers (Fig. S3). Independent of laser conditions, the maximum optical induced magnetization switching using a fixed layer beam is roughly half the saturation magnetization and is an accumulative effect using multiple optical pulses.

To probe the microstructural effect on the magnetization switching, we compare the Hall resistance change in the granular FePt-C film to a continuous 10-nm-thick FePt films (see methods and Fig. S1). While the continuous film has strong perpendicular anisotropy the perpendicular $H_c$ is only 0.17 T due to the small number of pinning site for magnetic domain wall motion in the epitaxial continuous film. Figure 2 shows the normalized Hall resistance change as a function of the RCP optical pulses in the FePt-C granular and FePt continuous films. The normalized Hall resistance decreases in both of the samples. However, in contrast to the granular FePt film, the continuous film simply demagnetizes with further increase of the RCP light pulse power (and this result is independent of laser helicity), which is consistent with the thermal demagnetization and the formation of domains as seen in thick Co/Pt multilayer films[11] driven by strong dipolar energies. In the granular film, the dipole energy is much smaller than that of the continuous film and the magnetization reversal cannot progress by domain wall motion because of the break in magnetic exchange between the magnetic grains. We have further shown a continuous transitional behavior between these two extreme behaviors in intermediate exchange composite granular/continuous films as shown in Fig. S4.

The data shown in Fig. 1(c) and (d) were fitted by a simple accumulative switching model where we treat the FePt grains as having two magnetic states, spin-up and spin-down state, because of its high magnetic anisotropy and small grain size. The total number of FePt grains N is expressed by the summation of spin-up grains $N\uparrow$ and spin-down grains $N\downarrow$ and the normalized magnetization is given by $(N\uparrow - N\downarrow)/(N\uparrow + N\downarrow)$. We assume that with each optical pulse there is a switching probability from the spin-up to spin-down state given by $P_1$ and from spin-down to spin-up by $P_2$. The number of FePt grains with spin-up and spin-down states after the *n* pulses can be expressed by the following:

$$N_\uparrow^n = N_\uparrow^{n-1}(1 - P_1) + N_\downarrow^{n-1}P_2 \quad \text{and} \quad N_\downarrow^n = N - N_\uparrow^n \tag{1}$$

The fitted lines to Eq. (1) are shown in the Fig. 1(c) and (d) using the same switching probabilities for RCP, LCP and linearly polarized light exposure: $(P_1, P_2)$ = (0.0048, 0.0016), (0.0016, 0.0048) and (0.0032, 0.0032), respectively. The fits suggest that the switching probability by a single pulse is very small, less than 1%. However, accumulating the small switching probabilities results in a continuous change in the magnetization until the final equilibrium state where $N\uparrow P_1 = N\downarrow P_2$ (a final state normalized magnetization of $(P_2-P_1)/(P_2+P_1)$ ) is reached. This final state is independent of the initial state of the magnetization as seen experimentally.

What is the origin for different switching probability of $P_1$ and $P_2$ for circularly polarized light pulses? The fact that linear light leads to demagnetization of the sample suggest that heating of the FePt grains by the femto-second laser exposure is sufficient to cause thermal activated reversal. The circularly polarized light then breaks the symmetry of the system favoring one magnetic state over the other and leading to an imbalance in the $P_1$ and $P_2$. This symmetry breaking could result from a direct interaction between the light and the magnetic systems such as the inverse Faraday field that prefers one direction over the other[22]. The difference in $P_1$ and $P_2$ could also arise from differential absorption for RCP and LCP (i.e. magnetic circular dichroism) that will result in a slight difference in temperature for one set of grains compared to the other.[17] We have roughly estimated the difference in temperature needed during the optical excitation to explain the difference in switching probability using a simple Arrhenius-Néel model for single domain particles (see supplementary section for details). A temperature difference of 1-2 K would be sufficient to explain the difference in $P_1$ and $P_2$ observed in Fig. 1, which is consistent with typical dichroism differences in magnetic metals.[17]

However, independent of the mechanism we find that magnetic switching for granular FePt films is statistical in nature in contrast to the reports on GdFeCo films. Because of this we don't achieve full deterministic switching, which would be needed for magnetic recording applications. To achieve full switching, we additionally applied a small external magnetic field, while applying 80 optical pulses at the laser power of 1.00 mW for each sequence. Figure 3(a) shows the normalized Hall resistance change after exposure to the linear polarized laser under an external magnetic field of 0.051 T. Regardless of the polarity of the external magnetic field, the magnetization of the FePt-C granular film remains close to the demagnetized state. When the external magnetic field is increased to ~0.2 T as shown in Fig. 3(b), the magnetization changes depending on the external magnetic field polarity, but only about half of the total magnetization can be switched by a 0.2-T field. However, by the combination of the circularly polarized light and an external magnetic field of 0.2 T, a magnetization reversal >90 % of the saturation magnetization can be achieved as shown in Fig. 3(c).

The results described here show that the AOS observed in FePt granular media can be described by a statistical model with a small probability of switching magnetic grains for each light pulse and these probabilities depend on the helicity of the light. This is qualitatively different from the original reports for GdFeCo, but closer to the behavior reported for Pt/Co/Pt structures[23]. We show that nearly deterministic switching can be achieved by the combination of circularly polarized light and small external fields suggesting that the control of the light helicity can aid the writing in a HAMR like recording process. This study shows that the fully deterministic switching in using only AOS for high-anisotropy nanostructures will require more complex structures.

.

**Methods**

**Film and device fabrication:**

The FePt-30vol%C (hereafter, FePt-C) granular film was deposited by co-sputtering of FePt and C targets on a MgO(001) single crystal substrate by DC magnetron sputtering under 0.48 Pa Ar pressure at 600°C. After the deposition of FePt-C, the sample was cooled down to RT and 10-nm-thick C was deposited as a capping layer. In this work, we used a compositionally graded sputtering method to obtain the columnar FePt grains[24]. In this process, the C composition gradually changed during the deposition, which gives nice FePt grain isolation as shown in Fig. S1(a). The FePt continuous film was deposited by the sputtering from an FePt alloy target under 0.48 Pa Ar pressure at 400°C. A 10-nm-thick C capping was deposited on the FePt continuous film at RT. The detailed description of the film processing conditions and their magnetic characteristics were reported elsewhere[25]. Hall crosses were used for the measurement of the magnetization change by the light exposure and the applied magnetic fields. They were fabricated directly from the films by photo-lithography using a lift-off process and a subsequent Ar ion-milling step. The width of the Hall cross is about 15 μm.

**Optical setup and microstructural, magnetic measurements**:

Figure S2 shows the optical setup for this experiment. A 590 nm wavelength LED was used for the observation of the magnetic domain of the sample. To excite the magnetization state, we use femtosecond laser exposures with center wavelength of 800 nm, the pulse duration of 150 fs and the repetition rate of 1 kHz. Femto-second laser pulses was focused onto a spot on the sample whose diameter is 20-50 μm, which covers the whole region of the Hall cross. To estimate the magnetization change by the light illumination, Hall cross devices were set at the focal point of the light. After exposing the sample to several optical pulses, the Hall voltage was measured. The number of pulses exposed on the sample was controlled by the shutter.

The sample microstructure and magnetic properties were measured by transmission electron microscopy (TEM) and vibrating sample magnetometer (VSM), respectively.

Acknowledgment

This work was supported in part by Grand-in-Aid for Scientific Research (B) Grant Number 26289232 and the work at UCSD was supported by the Office of Naval Research (ONR) MURI program.

Author contributions    Y.K.T and E.E.F designed and coordinated the projects; J.W, S.H.W., O.H. and K.H. grew and characterized the thin film samples. Y.K.T, R.M., S.K. and K.I performed the laser experiment. Y.K.T and E.E.F. coordinated work on the paper with contributions from R.M., S.K., J.W., K.I., S.H.W, O.H. K.H. with regular discussions with all authors.

Competing financial interests        The authors declare no competing financial interests.

**Fig. 1 Magnetization change observed from a FePt-C granular film by exposure to circular polarized light.** (a) Magnetic image after exposure to right and left circular polarized light at the laser power of 0.70 mW. The initial state was demagnetized. We show a subtracted image obtained from before and after applying the laser exposure. (b) Anomalous Hall Effect curve for the FePt-C granular film. The additional data points at zero field correspond to the amount of magnetization switched by exposure to right and left circular polarized light in the Hall cross as indicated. (c) Normalized Hall resistance after applying circular and linear polarized light as a function of the integrated number of pulses. Red, blue and black dots correspond to the normalized Hall resistance after applying the RCP, LCP and linear polarized light. The initial state is the spin-up state in all FePt grains. The dotted lines show the fitting results using the accumulative magnetic switching model described in the main text. The laser power was 1.28 mW. (d) Normalized Hall resistance after illumination with RCP, LCP and linear polarized light. The initial state is the spin-down state in all FePt grains.

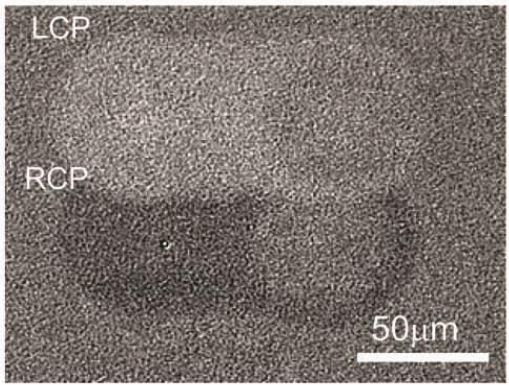
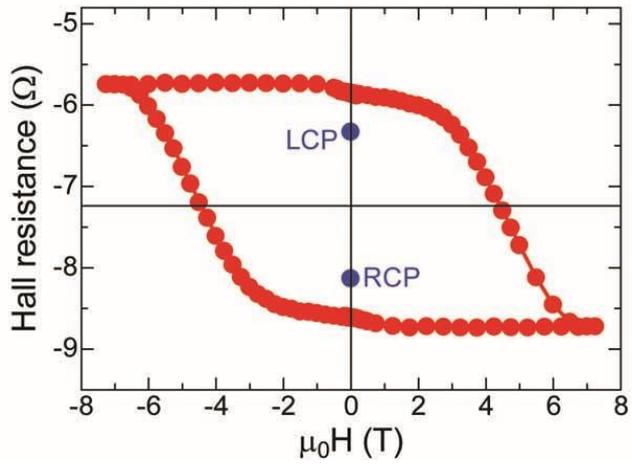
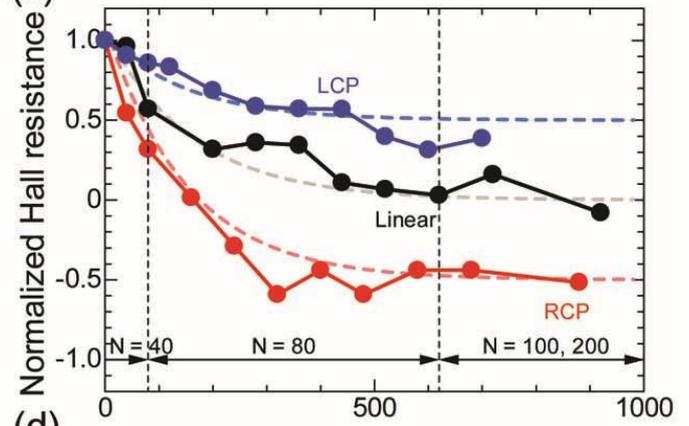
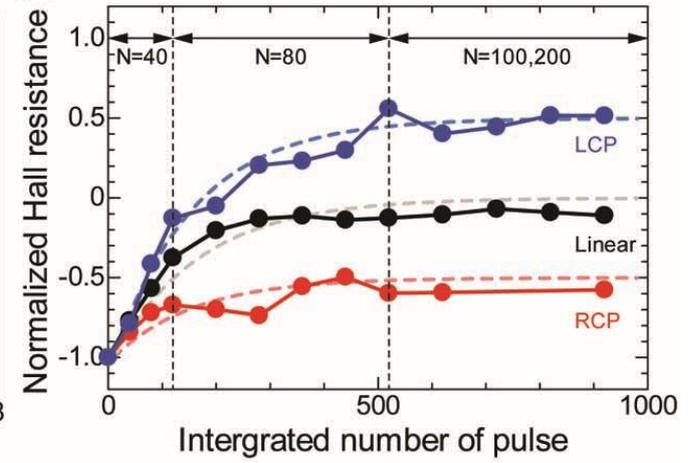

**Fig. 2 Normalized Hall resistance for FePt continuous (red) and FePt-C granular (blue) films versus light power.** The initial state was spin-up in FePt grains. The number of pulses was 80 for both samples. RCP light was used.

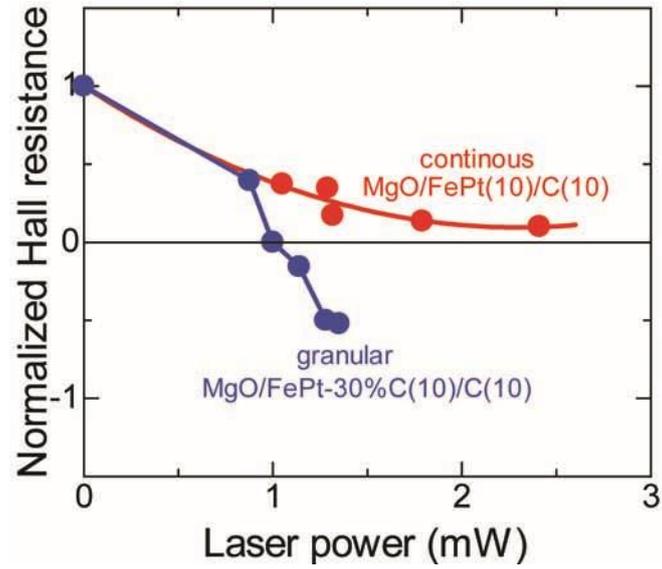

**Fig. 3 Deterministic switching by combination of circularly polarized light illumination with applying small magnetic field.** Normalized Hall resistances after the exposure to linear polarized light with (a) 0.051 T and (b) 0.2 T. The pulse number is 80 and the light power is 1.35 mW. (c) Normalized Hall resistance after the illumination with left and right circular polarized light and magnetic field of 0.2 T.

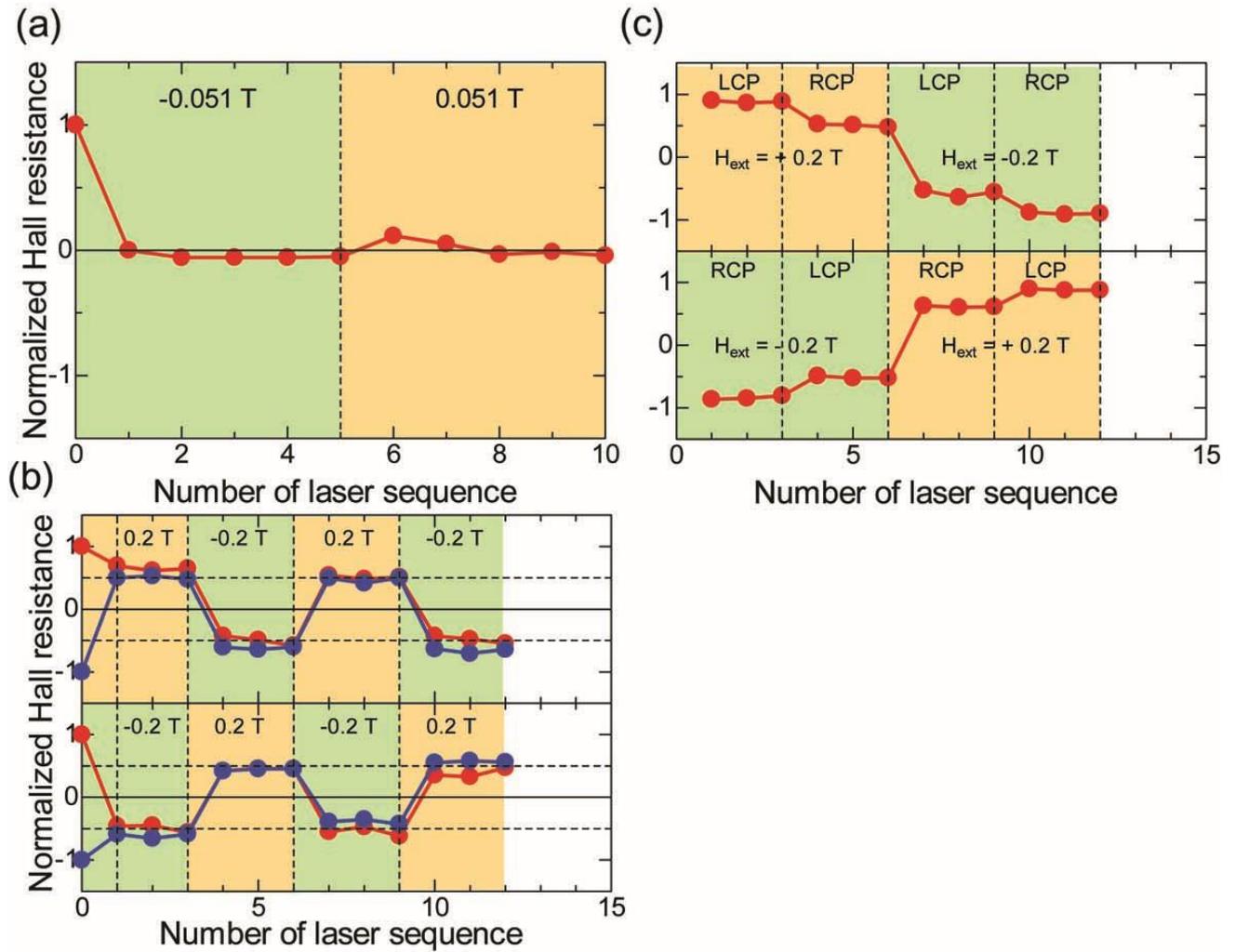

Supplementary information

# Accumulative magnetic switching of ultra-high-density recording media by circularly polarized light

Y.K. Takahashi, R. Medapalli, S. Kasai, J. Wang, K. Ishioka,

S.H. Wee, O. Hellwig, K. Hono and E.E. Fullerton

**Microstructure and magnetic properties of FePt-C and FePt films**

Figure S1(a) shows the transmission electron microscopy (TEM) bright field image of the FePt-C films along with the corresponding nanobeam diffraction (ND) pattern and histogram of the grain sizes. The dark contrast areas are the FePt grains and the bright contrast area is the carbon segregant. The average grain size is ~11.9 nm and these grains are distributed within an amorphous C matrix. The ND pattern indicates that all of the FePt grains have strong (001) texture because of the epitaxial growth on the MgO(001) single-crystal substrate. Figure S1(b) shows a cross sectional TEM image of the FePt-C granular film. The aspect ratio of the FePt grains is roughly 1. Figure S1(c) shows the magnetization curves of FePt-C granular film. Red and blue symbols correspond to the magnetization with the applied magnetic field perpendicular and parallel to the film surface, respectively. The film shows strong perpendicular magnetic anisotropy with a saturation field of 7 T and high coercivity field ($\mu_0H_c$) of 4 T due to the strong (001) texture and highly $L1_0$ ordered FePt grains. Figure S1(d) shows the corresponding TEM image and ND pattern of the FePt continuous film. The film shows a continuous microstructure and epitaxial growth on the MgO(001) substrate. In contrast to the magnetic properties of the granular film shown in Fig. S1(c), the $\mu_0H_c$ of the FePt continuous film is only 0.17 T due to the small number of pinning sites for the domain wall motion. Although the $\mu_0H_c$ is small, the anisotropy field is about 4 T. The smaller anisotropy field in FePt continuous films is due to the low degree of order (S=0.51) originated from lower deposition temperature of 400°C.

**Magnetization change as a function of laser power and number of pulses**

Figure S3 shows the change of the normalized Hall resistance of FePt-C film measured with increasing pulse number for various laser powers. The initial states is spin-up for all FePt grains and the sample was exposed to RCP light. At the lower power of 1.00 mW shown in Figure S3(a), exposing the sample to a series of 20 optical pulses does not significantly perturb the system. By increasing the number of pulses between measurements of the Hall voltage, the magnetization decreases. However, we do not observe a net switching into the opposite direction. This indicates that the small light power does not provide enough instantaneous heat for a significant magnetization switching to occur, but that only accumulated heat from many optical pulses at1 kHz rep rate can slowly thermally demagnetize the sample. At higher power of 1.35 mW shown in Fig. S3(b), the Hall resistance decreases much faster after the first laser exposure. In sets of 20 and 40 pulses, the Hall resistance decreases significantly by illumination of RCP light, however, the overall magnetization direction of the film still remains positive, just as for the initial state. When using sets of 60 and 80 pulses, the normalized Hall resistance also changes sign to -0.5. However, for sets of 160 pulses at a time, the normalized Hall resistance decreases only to -0.4, which may be due to excess accumulated heat in the sample from the longer sequence of pulses. Further increase of the pulse number such as 1000 pulses results in an irreversible changes to sample and the Hall cross response.

**Estimation of the temperature during the laser exposure**

The difference in $P_1$ and $P_2$ discussed in the main text could result from differential absorption resulting from the circular dichroism. This difference in absorption could lead to a slight difference in temperature for one set of grains compared to the other. To roughly estimate the temperature difference needed to explain the difference in switching probability, we use a simple Arrhenius-Néel model for single domain particles where the time-dependence of magnetization of an ensemble of particles versus time (t) is given by:

$$M(t) = M(t=0) \exp\left(\frac{-t}{\tau}\right) \qquad (2)$$

where $\tau$ is the characteristic time for thermal switching. In zero external field, $\tau$ is given by:

$$\tau = \tau_0 \exp\left(\frac{K_U V}{k_B T}\right) \qquad (3)$$

where $K_u V$ is the magnetic energy of the particle that is the product of the magnetic anisotropy $K_u$ and the particle volume V and $\tau_0$ is the attempt time for thermal activation (typically 0.1 ns for high-anisotropy magnetic systems). Assuming that M(t)/M(t=0) equals $1-P_1$ for RCP light and $1-P_2$ for LCP, then one can estimate $\tau_{LCP}/\tau_{RCP}$ assuming the time *t* is the same for RCP and LCP excitation by:

$$\frac{\tau_{LCP}}{\tau_{RCP}} = \frac{\ln(1-P_1)}{\ln(1-P_2)}. \qquad (4)$$

Combing Eq. (2) and Eq. (3), we can estimate the difference in $K_uV/k_BT$ between RCP and LCP excitations, which is given by

$$\left(\frac{K_UV}{k_BT}\right)_{LCP} - \left(\frac{K_UV}{k_BT}\right)_{RCP} = \ln\left(\frac{\tau_{LCP}}{\tau_{RCP}}\right) . \qquad (5)$$

For $P_1$= 0.0048 and $P_2$=0.0016, then Eq. (4) gives $\tau_{LCP}/\tau_{RCP}$ = 3.005 and the corresponding difference in $K_uV/k_BT$ between LCP and RCP excitations from Eq. (5) is 1.1.

From the properties of FePt grains, we estimate the temperature difference to result in a change in $K_uV/k_BT$ of 1.1. For a magnetic grain with 12 nm diameter, 10 nm height, and a magnetic anisotropy of $K_u$=4.3x10$^7$ ergs/cm$^3$, we can estimate room temperature $K_uV/k_BT$ is ~1200. Since $K_uV/k_BT$ goes to zero at $T_C$ ~ 700 K, we can roughly estimate that $K_uV/k_BT$ decreases on the order of 3 per K temperature change suggesting a small difference in temperature of 0.3 K due to excitation of RCP and RCP is consistent with the observed $P_1$ and $P_2$ values. However, these estimates ignore any interactions between particles, any distributions in particle size or the specific dependence of temperature on time after excitation.

We can further estimate how close to $T_C$ we would have to heat the sample to see the statistical switching observed. Using Eqs. (1) and (2) and assuming the sample is exposed for 1 ns, $K_uV/k_BT$ would need to be 7.6 to have a switching probably of $P_1$=0.0048 and 8.7 for the switching probability of $P_2$=0.0016, suggesting we are heating close to $T_C$.

In the case of the FePt continuous film, the absorbed heat from the light can be dispersed quickly. Therefore, the film is not irreversibly damage even at 2 mW. However, for the granular film, FePt grains are heated more effectively by the light exposure because of the thermal isolation of adjacent grains by the amorphous Carbon segregant whose thermal conductivity is more than ten times lower than that of FePt. For laser exposure higher than 1.5 mW, it results in irreversible damage of the FePt-C granular film.

**AOS in exchange-coupled composite (ECC) media**

Exchange-coupled composite (ECC) media was proposed in order to solve the so-called "trilemma problem" [1,2]. By putting the magnetically soft layer on top of the CoCrPt or FePt granular layer, the switching field can be effectively reduced while maintaining the energy barrier for the magnetization rotation[1-4]. Wang *et al*. reported the significant reduction of switching field in FePt-C granular media by adding a continuous FePt layer to the high anisotropy granular layer[5], which we use for the AOS experiment here. Figure S4 shows the schematic view of the film stack, magnetization curves and microstructures of the ECC media with various FePt layer thicknesses. The FePt layer was deposited at 400°C. Therefore, the degree of $L1_0$ ordering is lower than that of the FePt-C granular layer. The FePt-C granular layer consists of well-isolated FePt grains with uniform microstructure and average grain size of 10 nm. The morphology of the capping layer changed from partially granular to continuous when increasing the FePt cap layer thickness. With increasing the thickness of the FePt layer from 0 to 15nm, the perpendicular $H_c$ reduces from 4.9 T to 1.4 T. Figure S5 shows the change of the normalized Hall resistance of the films after exposure to RCP light across the Hall cross with a power of 0.48 mW. The initial magnetization state was demagnetized. The magnetization switching ratio decreases with increasing thickness of the FePt layer. The inset shows the magnetization changes by circular polarized light in FePt-C/FePt(15nm) ECC media. The magnetization switches depending on the helicity of the light and shows zero when linearly polarized light was used. The resistance change by the light illumination is reproducible. However, the magnetization switching ratio is very small, only 0.3 %. These results also support that the lack of temperature and the large dipole energy only allow for a small magnetization switching in the continuous film.

**Fig. S1 Microstructure and magnetic properties of FePt-C and FePt films.** (a) plan-view TEM bright field image, corresponding nanobeam diffraction pattern and histogram of FePt grain size, (b) a cross sectional TEM image and (c) magnetization curves of FePt-C granular film. (d), (e) and (f) is the same data for the corresponding FePt continuous film.

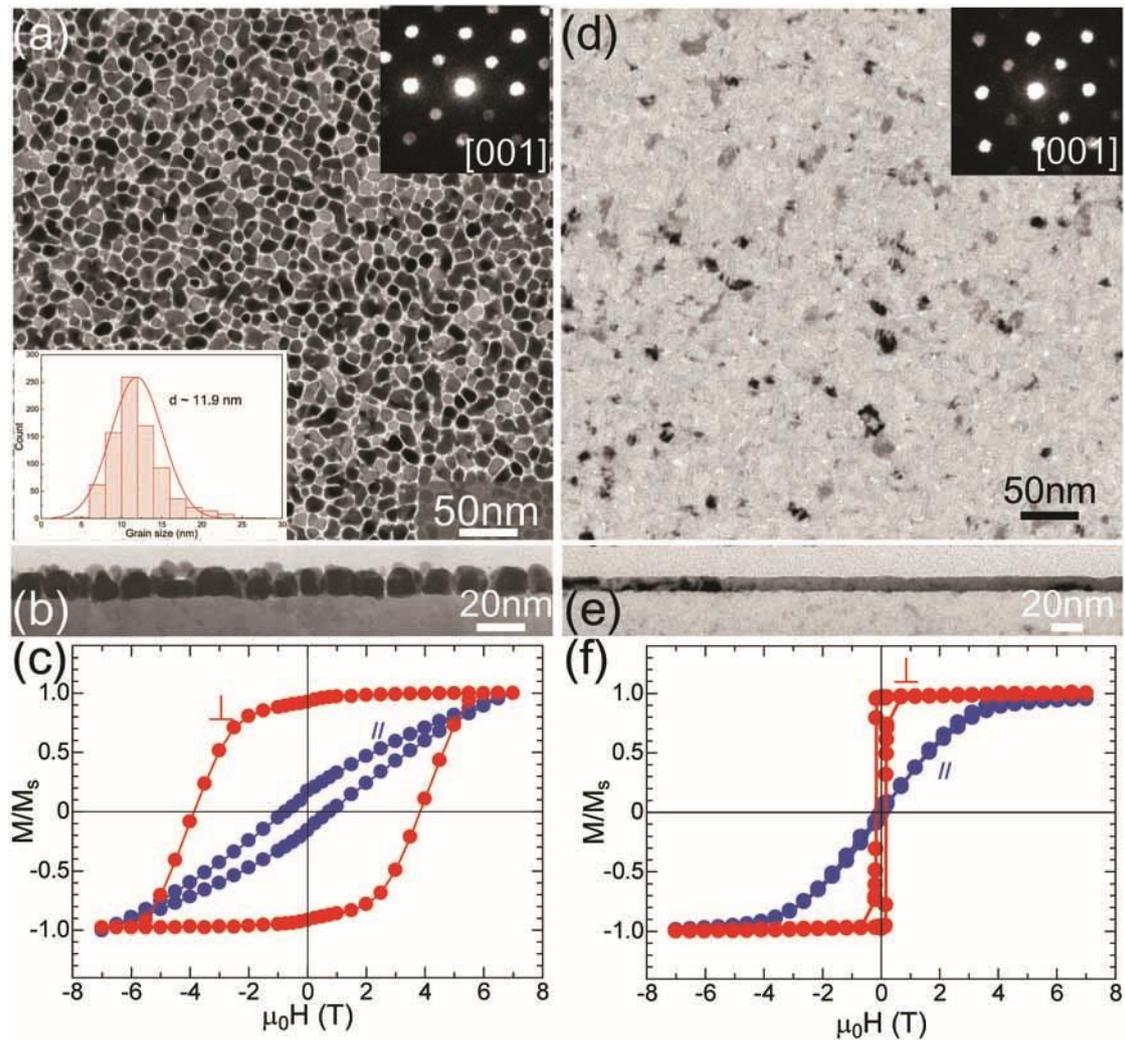

**Fig. S2 schematic view of the optical setup.** In the experiments for estimating the magnetization switching ratio, several pulses are illuminated right after each other on the Hall cross. The number of the pulses was controlled by the shutter. When we scan the laser on the sample, the sample was manually moved.

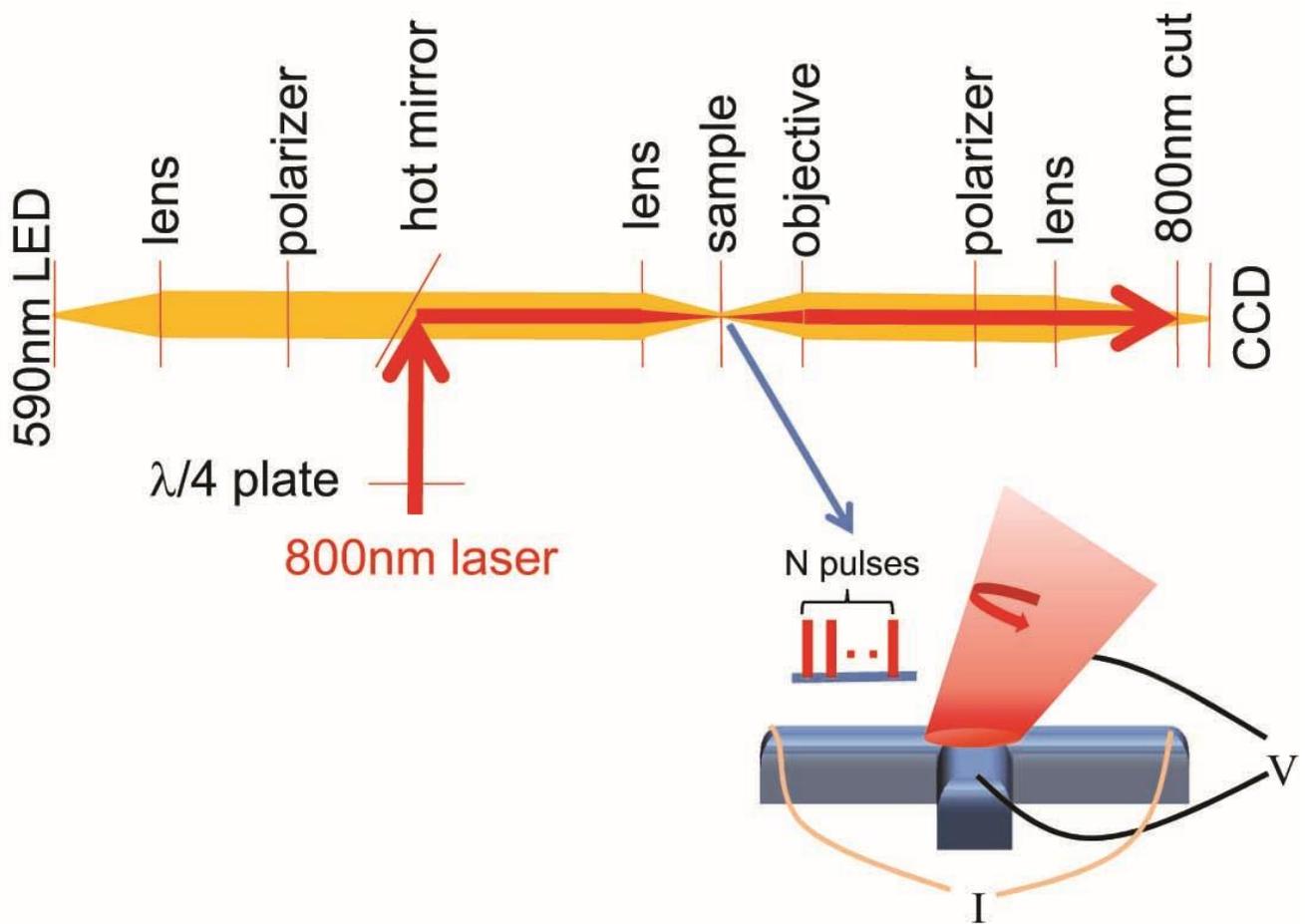

**Fig. S3 Magnetization change for FePt-C film at various conditions of right circular polarized light.** The number of optical pulses between Hall voltage measurements was varied from 20 to 160. The power of the light in (a) and (b) is 1.00 mW and 1.35 mW, respectively.

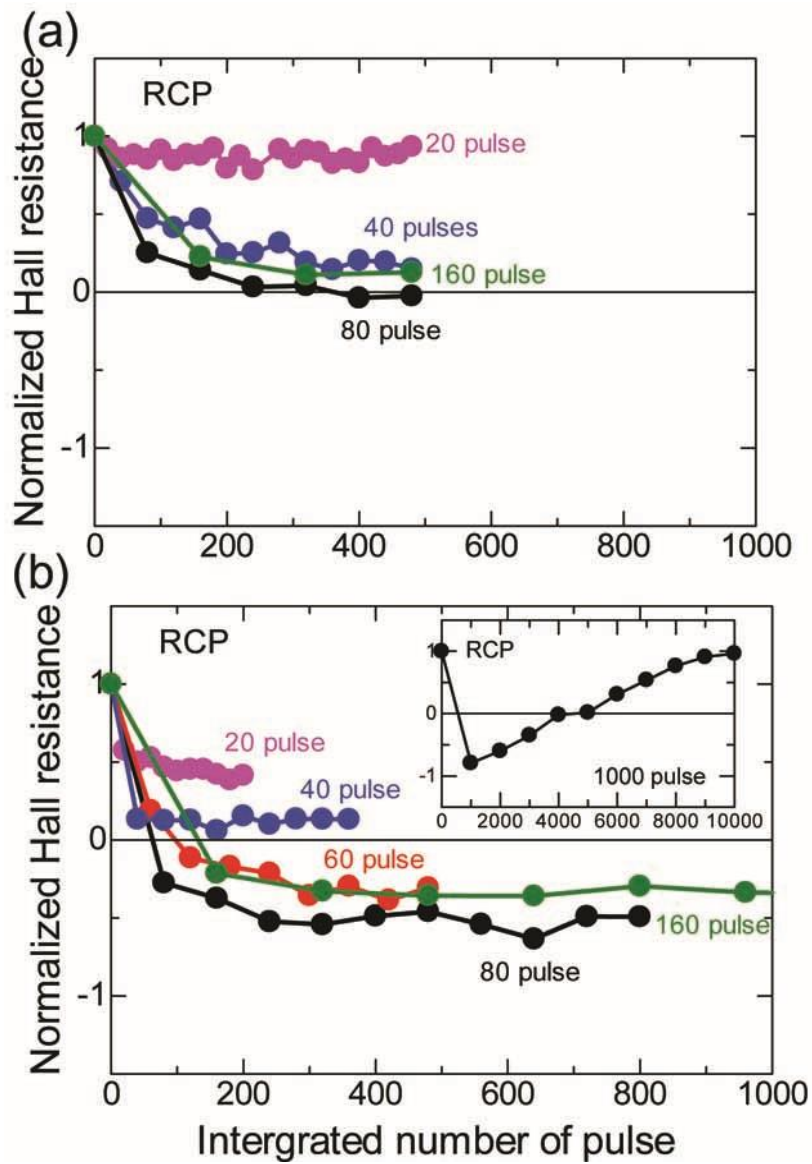

**Fig. S4 FePt-C ECC media.** Schematic view of the film stacks, magnetization curves and microstructures of FePt-C ECC media with various thicknesses of the semi-hard FePt layer[5].

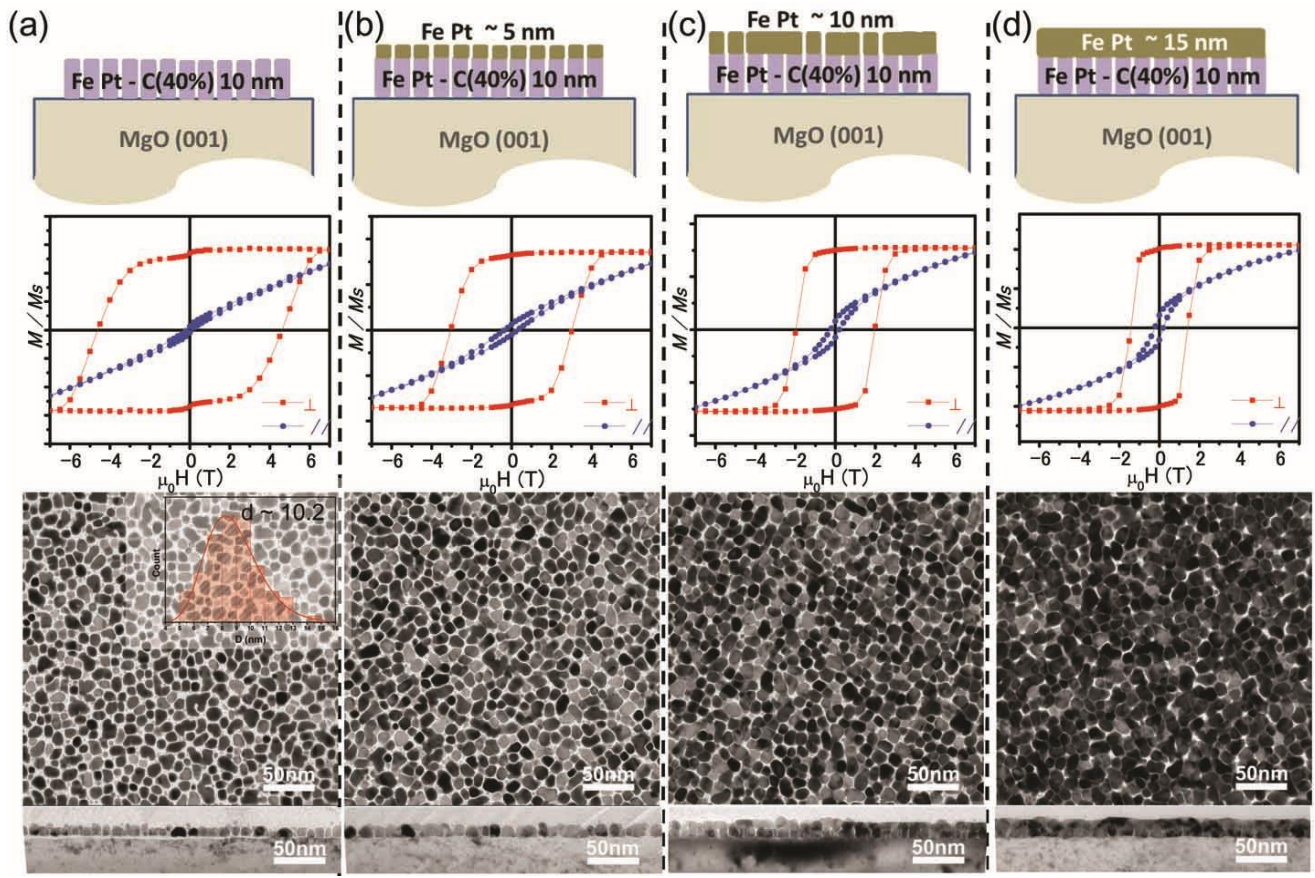

**Fig. S5 Magnetization switching in FePt-C ECC media.** The change of the magnetization switching ratio as a function of the film thickness of semi-hard FePt layer. The inset shows the magnetization change by the light illumination in FePt-C/FePt(15) ECC media.

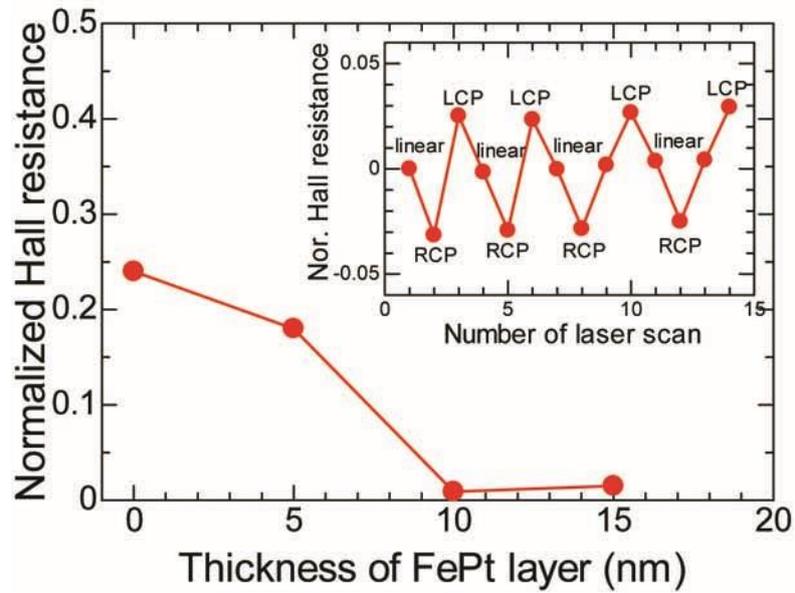